# Theoretical correction methods for optical tweezers: Acquisition of potentials of mean forces between colloidal particles in a bulk and on a surface

**Ken-ichi Amano,[a*] Rikako Suzuki,[a] and Madoka Takasu[a]**

[a] *Faculty of Agriculture, Meijo University, Nagoya, Aichi 468-8502, Japan.*

* Correspondence author: K. Amano (amanok@meijo-u.ac.jp)

**ABSTRACT:** It is known that line optical tweezers (LOT) can measure potential of mean force (PMF) between colloidal particles in the bulk. However, PMF obtained with LOT is empirically modified before showing the result of the final form in order to correct the potential rise at long distances. In the present letter, we derive theoretical correction methods for acquisition of PMF by using statistical mechanics. Using the new methods, PMF can be obtained without the empirical fitting equation. Through the new methods, external potential acting on the trapped two colloidal particles induced by LOT can also be obtained. As an additional study, we explain two methods for obtaining PMF between colloidal particles on a substrate surface, in which a normal single optical tweezers with a fixed focal point is used, and for obtaining PMF between colloidal particles trapped by dual-beam optical tweezers in the bulk. These methods can also obtain the external potential acting on the trapped two colloidal particles.

## MAIN TEXT

Line optical tweezers (LOT) can measure potential of mean force (PMF) between two colloidal particles in the bulk [1][2]. LOT has been used for studies of the colloidal interactions in various conditions [3][4]. It has contributed to a study of mechanism of dispersion stability of the colloidal particles.

LOT traps two colloidal particles within its scan line, and a sufficiently long time video of the trapped particles is taken with the video microscopy. From the video, a probability density function ($p$) in terms of particle-particle distance (distance between centers of the colloidal particles: $r$) is obtained by using an arbitrary particle tracking method. After obtaining $p(r)$ from the particle tracking method, PMF between colloidal particles in the bulk ($u(r)$) is calculated by using a following equation:

$$u(r) = -\frac{1}{\beta}\ln p(r) + C_0, \qquad (1)$$

where $\beta = 1/(k_B T)$, $k_B$ is the Boltzmann constant, $T$ is the absolute temperature, and $C_0$ is a constant (offset). Eq. (1) is derived from the following basic formula in statistical mechanics:

$$p(r) = \frac{\exp\left(-\beta u(r)\right)}{Z}, \qquad (2)$$

where $Z$ is the partition function. Generally, Eq. (1) is used for calculation of $u(r)$, but the obtained $u(r)$ has an arch shape. For that reason, it requires an empirical modification after use of Eq. (1) [1][2]. Through development and study of LOT, we also obtained $u(r)$ with the arch shape. We think that the empirical modification is one of the important approaches to obtain plausible $u(r)$, that is, we do not deny the empirical modification. However, the empirical modification does not have sound theoretical back up. Although the empirical modification can dissolve the potential rise at long distances, it is not clear whether or not the modifications of the short- and medium-distances are correct. Therefore, in the present letter, we create theoretical correction methods for acquisition of the proper $u(r)$ based on statistical mechanics.

In our experiment, an existence probability distribution of a colloidal particle in the scan line did not become equal probability. In most cases, the trapped particle moved to only one side of the scan line and Brownian motion was observed in the limited region. The same was true when two colloidal particles were trapped. Here, we mention the region that Brownian motions were observed as a sub-scan line, and set the width of the sub-scan line from $-l$ to $+l$. Assuming that external potential is formed on the sub-scan line and two particles undergo Brownian motion for a sufficiently long time in the sub-scan line, the following partition function can be written:

$$Z = \int_{-l}^{l} \int_{-l}^{l} \exp(-\beta[u(r) + v(x_1) + v(x_2)]) dx_1 dx_2, \qquad (3)$$

where $v(x_i)$ is the external potential acting on the colloidal particle placed at $x_i$. Eq. (3) is a conventional form of the partition function for that experimental situation. However, Eq. (3) is not sufficiently correct for the particles video taken. In the video, positions of particle 1 and particle 2 are not swapped, because the scan line is one dimensional (1D). Thus, the partition function must be formulated under the following condition: $x_1 < x_2$ or $x_1 \leq x_2$. Furthermore, the partition function must be constructed by using $r$ as the leading variable so that one can link $p(r)$ and $u(r)$ with an easy-to-use form. We newly express such a partition function as $Z_r$:

$$Z_r = \int_{0}^{2l} f(r) dr, \qquad (4)$$

where $f(r)$ is related to $p(r)$,

$$p(r) = \frac{f(r)}{Z_r}. \qquad (5)$$

Here, $p(r)$ is the probability density function that becomes 1 when integration with respect to $r$ is performed from 0 to $2l$. The function $f(r)$ is given by

$$f(r) = \int_{-l}^{l-r} \exp(-\beta[u(r) + v(x_1) + v(x_1 + r)])dx_1. \qquad (6)$$

With that partition function, statistics of the two colloidal particles in the sub-scan line is correctly expressed.

In what follows, we derive an equation for calculation of $u(r)$ by using Eqs. (4)-(6). Taking natural logarithm, Eq. (5) is rewritten as

$$\ln p(r) = \ln f(r) - \ln Z_r. \qquad (7)$$

Substitution of Eq. (6) into the first term on the right-hand side of Eq. (7) gives

$$\begin{aligned} \ln f(r) &= \ln\left(e^{-\beta u(r)} \int_{-l}^{l-r} e^{-\beta(v(x_1)+v(x_1+r))} dx_1\right) \\ &= -\beta u(r) + \ln \int_{-l}^{l-r} e^{-\beta\left(v(x_1)+v(x_1+r)\right)} dx_1. \qquad (8) \end{aligned}$$

Thus, $u(r)$ is expressed as

$$u(\mathrm{r}) = -\frac{1}{\beta}\ln p(r) + \frac{1}{\beta}\ln\int_{-l}^{l-r} e^{-\beta[v(x_1)+v(x_1+r)]}dx_1 + C_1, \qquad (9)$$

where $C_1$ ($= -(1/\beta)\ln Z_r$ ) is a constant (offset). Appearance of the integral term in Eq. (9) is a new term in comparison with Eq. (1). Next, we calculate the integral term by substituting a well-type potential into $v(x_i)$. We assume that $v(x_i) = -\varepsilon/\beta$ when $-l \leq x_i \leq +l$ ($\varepsilon$ is positive value). In that case, $e^{-\beta[v(x_1)+v(x_1+r)]} = e^{2\varepsilon}$, and hence the integration can be readily performed. Consequently, $u(r)$ is expressed as

$$u(r) = -\frac{1}{\beta}\ln p(r) + \frac{1}{\beta}\ln(2l - r) + C_2, \qquad (10)$$

where $C_2$ ($= C_1 + 2\varepsilon/\beta$) is a constant (offset). If the external potential is well-type potential, $u(r)$ is calculated by using both Eq. (10) and $p(r)$ obtained from the video. For calculation of $u(r)$, it is necessary to input appropriate $l$ and $C_2$. Appropriate $l$ flattens $u(r)$ at the long distances and appropriate $C_2$ converges $u(r)$ at the long distances to 0. If inputted $l$ is proper, the value of $l$ is close to supposed width of the sub-scan line from the video images. The parameters $l$ and $C_2$ can be uniquely determined in the calculation of $u(r)$ simultaneously. We name this method as 1D well-type potential resuscitation (WPR) method. In the 1D-WPR method, Arrhenius plot like technique can be used for determination of $l$ and $C_2$. Since 1D-WPR method can obtain not only the PMF but also the width of the sub-scan line $l$, this method can be called an inverse analysis method for determination of the width. By the way, the conventional empirical correction method of $u(r)$ has uncertainty especially at the short- and medium-distances. However, the 1D-WPR method does not have such an uncertainty if the external potential in the sub-scan line is (almost) well-type.

Second, we explain another correction method by assuming parabolic-type external potential. We name it as 1D parabolic-type potential resuscitation (PPR) method. Substituting $v(x_i) = ax_i^2 - \varepsilon/\beta$ into Eq. (9), we obtain

$$u(r) = -\frac{1}{\beta}\ln p(r) + \frac{1}{\beta}\ln \int_{-l}^{l-r} e^{-\beta a\left(2x_1^2+2x_1 r+r^2\right)}\, dx_1 + C_2. \qquad (11)$$

The exponential term in Eq. (11) is rewritten as

$$e^{-2a\beta\left(x_1^2+x_1 r+\frac{1}{2}r^2\right)} = e^{-2a\beta\left\{\left(x_1+\frac{r}{2}\right)^2+\frac{1}{4}r^2\right\}} = e^{-2a\beta w^2}\cdot e^{-\frac{a\beta r^2}{2}}, \qquad (12)$$

where we used the following definition $x_1 + r/2 \equiv w$. Substituting Eq. (12) into the integral term ($\equiv g(r;\, a,\, l)$) in Eq. (11), we obtain

$$g(r;\, a, l) = e^{-\frac{a\beta r^2}{2}} \int_{-l+\frac{r}{2}}^{l-\frac{r}{2}} e^{-2a\beta w^2} dw. \qquad (13)$$

Here, we used following equations: $2a\beta w^2 = \left(\sqrt{2a\beta}\cdot w\right)^2$, $\sqrt{2a\beta}\cdot w \equiv s$, and $ds/dw = \sqrt{2a\beta}$. Then, $g(r;\, a,\, l)$ can be rewritten as

$$g(r;\, a, l) = \frac{e^{\frac{-a\beta r^2}{2}}}{\sqrt{2a\beta}} \int_{\left(-l+\frac{r}{2}\right)\sqrt{2a\beta}}^{\left(l-\frac{r}{2}\right)\sqrt{2a\beta}} e^{-s^2} ds$$

$$= \frac{e^{-\frac{a\beta r^2}{2}}}{\sqrt{2a\beta}} \left\{ \int_{\left(-l+\frac{r}{2}\right)\sqrt{2a\beta}}^{0} e^{-s^2} ds + \int_{0}^{\left(l-\frac{r}{2}\right)\sqrt{2a\beta}} e^{-s^2} ds \right\}$$

$$= \frac{e^{-\frac{a\beta r^2}{2}}}{\sqrt{2a\beta}} \cdot \frac{\sqrt{\pi}}{2} \left\{ \operatorname{erf}\left(\left(l-\frac{r}{2}\right)\sqrt{2a\beta}\right) - \operatorname{erf}\left(\left(\frac{r}{2}-l\right)\sqrt{2a\beta}\right) \right\}, \qquad (14)$$

where erf represents the error function. The error function is generally expressed as

follows:

$$\int_0^x e^{-t^2} dt = \text{erf}(x)\frac{\sqrt{\pi}}{2}. \qquad (15)$$

Substitution of Eq. (14) into Eq. (11) yields

$$u(r) = -\frac{1}{\beta}\ln p(r) + \frac{1}{\beta}\ln\left(\frac{e^{-\frac{a\beta r^2}{2}}}{\sqrt{2a\beta}}\cdot\frac{\sqrt{\pi}}{2}\left\{\text{erf}\left(\left(l-\frac{r}{2}\right)\sqrt{2a\beta}\right)-\text{erf}\left(\left(\frac{r}{2}-l\right)\sqrt{2a\beta}\right)\right\}\right) + C_2. \qquad (16)$$

Eq. (16) can be used for acquisition of $u(r)$. By the way, if the external potential at $\pm l$ (i.e., $v(\pm l)$) is sufficiently high compared with that at center (i.e., $v(0)$), one can acquire $u(r)$ in an easier way. This situation enables us to substitute $l = \infty$ into Eq. (16), and then we obtain

$$u(r) = -\frac{1}{\beta}\ln p(r) - \frac{ar^2}{2} + C_3, \qquad (17)$$

where $C_3$ ($= C_2 + (1/(2\beta))*\ln(\pi/(2a\beta))$) is a constant (offset). $C_3$ contains $a$ and $\beta$ but it can be regarded as a simple constant when $u(r)$ is calculated from $p(r)$. Through calculation of $u(r)$, the parabola parameter $a$ is also obtained. That is, the 1D-PPR method can be used as an inverse analysis method of shape of the external potential in the sub-scan line. When two colloidal particles closely exist in the sub-scan line, trapping laser light is reflected and scattered from the colloidal particles. The reflected and scattered lights affect the stability of the neighbor particle, which changes the shape of the external potential. Such a nonlinear optical effect is difficult to predict in theory, because the inner structure of the colloidal particle and its surface structure including electric double layer are very complicated. Therefore, calculation of the parabola parameter $a$ is one of the important approaches for study of the nonlinear optical effect.

We tested the 1D-WPR and 1D-PPR methods by using $p(r)$ obtained from our LOT. Although it is now shown, we found that the 1D-PPR method is better than the 1D-WPR method in quality. This is because the 1D-PPR method could properly flatten the PMF at the long distances compared with the 1D-WPR method. From the result, we conclude that shape of the external potential in the sub-scan line is parabolic type rather than well type.

Third, we derive a two-dimensional (2D) method for acquisition of $u(r)$. We set a following experimental condition: (I) position of the focal point of the laser light is fixed; (II) the focal point exists near a substrate surface; (III) the shape of the external potential induced by the laser light is 2D parabolic type; (IV) two colloidal particles are trapped within the external potential; (V) The trapped colloidal particles do not change heights from the substrate surface. In this study, the method for acquisition of $u(r)$ is named 2D-PPR method. In that condition, conventional partition function in polar coordinates is given by

$$Z = \int_0^R dr_2 \, r_2 \int_0^R dr_1 \, r_1 \int_0^{2\pi} d\theta_2 \int_0^{2\pi} d\theta_1 \exp(-\beta\{u(r) + v(r_1) + v(r_2)\}), \qquad (18)$$

where $r$ represents distance between the centers of the colloidal particles. In Eq. (18), $r$ is calculated as follows:

$$r = \sqrt{(r_2\cos\theta_2 - r_1\cos\theta_1)^2 + (r_2\sin\theta_2 - r_1\sin\theta_1)^2}. \qquad (19)$$

We would like to obtain $u(r)$ from $p(r)$, but Eq. (18) does not have a formula that clearly related to $p(r)$. To obtain $u(r)$ from $p(r)$, we create new polar coordinates for two colloidal particles (Fig. 1). Under the polar coordinates, the partition function ($Z_r$) is given by

$$Z_\mathrm{r} = \int_0^{2R} f(r)dr. \qquad (20)$$

In that form, the probability density function $p(r)$ can be obtained by using a basic formula in statistical mechanics:

$$p(r) = \frac{f(r)}{Z_r}, \qquad (21)$$

where $f(r)$ is a function we must derive here (the concrete form is not known in this stage). In Eq. (21), $p(r)$ becomes 1 when integration with respect to $r$ is performed from 0 to $2R$. Here, we express $r_2$ in relation to $r_1$, $r$, and $\theta_2$ as follows:

$$r_2 = \sqrt{(r_1 + r\cos\theta_2)^2 + (r\sin\theta_2)^2}. \qquad (22)$$

When the circle which related to $\theta_2$ (see Fig. 1) has intersections, there must exist $\theta_2$ such that $r_2 = R$. When the number of the intersection is 2, they (defined as $\theta_{2\mathrm{a}}$ and $\theta_{2\mathrm{b}}$) are expressed as follows:

$$\theta_{2\mathrm{a}} = \arccos\left(\frac{R^2 - r_1^2 - r^2}{2rr_1}\right), \qquad (23)$$

$$\theta_{2\mathrm{b}} = 2\pi - \arccos\left(\frac{R^2 - r_1^2 - r^2}{2rr_1}\right). \qquad (24)$$

Considering two cases that the circle related to $\theta_2$ has no intersection and has two intersections, integration in the polar coordinates is written as

$$\int_0^R r_1 dr_1 \int_0^{R-r_1} drr \int_0^{2\pi} d\theta_2 \int_0^{2\pi} d\theta_1 + \int_0^R dr_1 r_1 \int_{R-r_1}^{R+r_1} drr \int_{\theta_{2\mathrm{a}}}^{\theta_{2\mathrm{b}}} d\theta_2 \int_0^{2\pi} d\theta_1 = (\pi R^2)^2. \qquad (25)$$

The integral with respect to $\theta_1$ is readily performed as follows:

$$2\pi \int_0^R r_1 dr_1 \int_0^{R-r_1} drr \int_0^{2\pi} d\theta_2 + 2\pi \int_0^R dr_1 r_1 \int_{R-r_1}^{R+r_1} drr \int_{\theta_{2\mathrm{a}}}^{\theta_{2\mathrm{b}}} d\theta_2 = (\pi R^2)^2. \qquad (26)$$

This integral form is not directly related to Eqs. (20) and (21). Hence, we alternate the integration order of $r$ and $r_1$:

$$2\pi \int_0^R \int_0^{R-r} \int_0^{2\pi} d\theta_2\, r_1 dr_1 r dr + 2\pi \int_0^R \int_{R-r}^{R} \int_{\theta_{2\mathrm{a}}}^{\theta_{2\mathrm{b}}} d\theta_2\, r_1 dr_1 r dr + 2\pi \int_R^{2R} \int_{r-R}^{R} \int_{\theta_{2\mathrm{a}}}^{\theta_{2\mathrm{b}}} d\theta_2\, r_1 dr_1 r dr = (\pi R^2)^2. \qquad (27)$$

The left-hand side of Eq. (27) is rewritten as

$$\begin{aligned} &2\pi \int_0^R \left[ \int_0^{R-r} \int_0^{2\pi} d\theta_2\, r_1 dr_1 + \int_{R-r}^{R} \int_{\theta_{2\mathrm{a}}}^{\theta_{2\mathrm{b}}} d\theta_2\, r_1 dr_1 \right] r\, dr + 2\pi \int_R^{2R} \left[ \int_{r-R}^{R} \int_{\theta_{2\mathrm{a}}}^{\theta_{2\mathrm{b}}} d\theta_2 r_1 dr_1 \right] r dr \\ &= 2\pi \int_0^{2R} \left[ \left( \int_0^{R-r} \int_0^{2\pi} d\theta_2 r_1 dr_1 + \int_{R-r}^{R} \int_{\theta_{2\mathrm{a}}}^{\theta_{2\mathrm{b}}} d\theta_2 r_1 dr_1 \right) H_0(R-r) + \left( \int_{r-R}^{R} \int_{\theta_{2\mathrm{a}}}^{\theta_{2\mathrm{b}}} d\theta_2 r_1 dr_1 \right) H_0(r-R) \right] r dr. \end{aligned} \qquad (28)$$

Here, $H_0$ is the Heaviside step function containing a property $H_0(0) = 0$. Before insertion of the Boltzmann factors into the integrals, we consider the external potential. That of parabolic-type in the polar coordinates can be written as $v(r_i) = ar_i^2 - \varepsilon/\beta$. Thus, sum of the external potentials acting on the two colloidal particles is

$$v(r_1) + v(r_2) = \mathrm{a}r_1^2 - \varepsilon/\beta + a[(r_1 + r\cos\theta_2)^2 + (r\sin\theta_2)^2] - \varepsilon/\beta$$
$$= 2a(r_1^2 + r_1 r\cos\theta_2 + r^2/2) - \frac{2\varepsilon}{\beta}. \qquad (29)$$

Since $u(r)$ and $-2\varepsilon/\beta$ can be placed outside of the integrals with respect to $\theta_2$ and $r_1$, the partition function can be expressed as

$$Z_r = 2\pi \int_0^{2R} drre^{-\beta u(r)+2\varepsilon}\left[\left\{\int_0^{R-r} dr_1 r_1 \int_0^{2\pi} d\theta_2 e^{-2a\beta\left(r_1^2+r_1 r\cos\theta_2+r^2/2\right)}\right\} H_0(R-r)\right.$$
$$+\left\{\int_{R-r}^{R} dr_1 r_1 \int_{\theta_{2\mathrm{a}}}^{\theta_{2\mathrm{b}}} d\theta_2 e^{-2a\beta\left(r_1^2+r_1 r\cos\theta_2+r^2/2\right)}\right\} H_0(R-r)$$
$$\left.+\left\{\int_{r-R}^{R} dr_1 r_1 \int_{\theta_{2\mathrm{a}}}^{\theta_{2\mathrm{b}}} d\theta_2 e^{-2a\beta\left(r_1^2+r_1 r\cos\theta_2+r^2/2\right)}\right\} H_0(r-R)\right]. \qquad (30)$$

Comparing Eq. (20) and Eq. (30), we can yield the concrete form of $f(r)$ as follows:

$$f(r) = 2\pi re^{-\beta u(r)+2\varepsilon}\left[\left\{\int_0^{R-r} dr_1 r_1 \int_0^{2\pi} d\theta_2 e^{-2a\beta\left(r_1^2+r_1 r\cos\theta_2+r^2/2\right)}\right\} H_0(R-r)\right.$$
$$+\left\{\int_{R-r}^{R} dr_1 r_1 \int_{\theta_{2\mathrm{a}}}^{\theta_{2\mathrm{b}}} d\theta_2 e^{-2a\beta\left(r_1^2+r_1 r\cos\theta_2+r^2/2\right)}\right\} H_0(R-r)$$
$$\left.+\left\{\int_{r-R}^{R} dr_1 r_1 \int_{\theta_{2\mathrm{a}}}^{\theta_{2\mathrm{b}}} d\theta_2 e^{-2a\beta\left(r_1^2+r_1 r\cos\theta_2+r^2/2\right)}\right\} H_0(r-R)\right]. \qquad (31)$$

Here, we define $g(r; a, R)$ as follows:

$$g(r;\ a,R) \equiv \left\{\int_{0}^{R-r} dr_1 r_1 \int_{0}^{2\pi} d\theta_2 e^{-2a\beta\left(r_1^2+r_1 r\cos\theta_2+\frac{r^2}{2}\right)}\right\} H_0(R-r)$$
$$+\left\{\int_{R-r}^{R} dr_1 r_1 \int_{\theta_{2\mathrm{a}}}^{\theta_{2\mathrm{b}}} d\theta_2 e^{-2a\beta\left(r_1^2+r_1 r\cos\theta_2+\frac{r^2}{2}\right)}\right\} H_0(R-r)$$
$$+\left\{\int_{r-R}^{R} dr_1 r_1 \int_{\theta_{2\mathrm{a}}}^{\theta_{2\mathrm{b}}} d\theta_2 e^{-2a\beta\left(r_1^2+r_1 r\cos\theta_2+\frac{r^2}{2}\right)}\right\} H_0(r-R). \quad (32)$$

Substitutions of Eqs. (31) and (32) into Eq. (21) yields

$$\ln p(r) = \ln f(r) - \ln Z_{\mathrm{r}} = -\beta u(r) + \ln(rg(r;a,R)) + \ln(2\pi) + 2\varepsilon - \ln Z_{\mathrm{r}}. \quad (33)$$

Consequently, $u(r)$ is expressed as follows:

$$u(r) = -\frac{1}{\beta}\ln p(r) + \frac{1}{\beta}\ln\big(rg(r;a,R)\big) + C_4, \quad (34)$$

where $C_4$ $(= 2\varepsilon/\beta + (1/\beta)*\ln(2\pi/Z_{\mathrm{r}}))$ is a constant (offset). Using Eq. (34), it is possible to caluclate $u(r)$ from $p(r)$. However, integrations in $g(r;\ a,\ R)$ is somewhat cumbersome, because it requires the numerical double integrations. Hence, we search a simple expression of $u(r)$ in the limit of $R\to\infty$. This search should provide a way to calculate $u(r)$ without the numerical double integrations. First, we calculate $Z_r$ in the limit of $R\to\infty$,

$$\lim_{R\to\infty} Z_r = 2\pi\int_{0}^{\infty} drre^{-\beta u(r)+2\varepsilon}\int_{0}^{\infty} dr_1 r_1 \int_{0}^{2\pi} d\theta_2 e^{-2a\beta\left(r_1^2+r_1 r\cos\theta_2+r^2/2\right)}. \quad (35)$$

Thus, $f(r)$ in the limit of $R\to\infty$ is

$$\lim_{R\to\infty} f(r) = 2\pi re^{-\beta u(r)+2\varepsilon}\int_{0}^{\infty} dr_1 r_1 \int_{0}^{2\pi} d\theta_2 e^{-2a\beta\left(r_1^2+r_1 r\cos\theta_2+r^2/2\right)}. \quad (36)$$

Performing the integrations, we obtain

$$\lim_{R\to\infty} f(r) = 2\pi r e^{-\beta u(r)+2\varepsilon}\left[\pi e^{-a\beta r^2/2}/(2a\beta)\right]. \qquad (37)$$

As a result, $u(r)$ can be calculated in a simple way in the limit of $R\to\infty$:

$$u(r) = -\frac{1}{\beta}\ln p(r) + \frac{1}{\beta}\ln(r) - \frac{ar^2}{2} + C_5, \qquad (38)$$

where $C_5$ ($= (1/\beta)\ln(\pi^2/(a\beta Z_r)) + 2\varepsilon/\beta$) is a constant (offset). Although it contains several variables, one can treat it as the constant in the calculation of $u(r)$.

Fourth, we shortly explain the resuscitation method for dual-beam optical tweezers[5] by introducing 2D trapping plane with two parabolic-type potentials. A following experimental condition is used for acquisition of the PMF ($u(r)$) between the trapped particles: (I) positions of the focal point of the laser lights are fixed; (II) the focal points exit in the bulk on the same plane; (III) the shapes of the external potential induced by the laser lights are 2D parabolic type; (IV) two colloidal particles are trapped within the external potential; (V) The trapped colloidal particles do not deviate from the plane. In this study, the method for acquisition of $u(r)$ is named 2D-DPPR method, where DPPR represents "dual-beam parabolic-type potential resuscitation". In that condition, the probability density function can be written with the same form of Eq. (21). In a specific polar coordinate for the condition of the dual-beam optical tweezers (see Fig. 2), the partition function $Z_r$ can be written as

$$Z_r = \int_0^{l_\mathrm{t}+r_{\mathrm{t}1}+r_{\mathrm{t}2}} f(r)\, dr, \qquad (39)$$

where $l_t$, $r_{t1}$, and $r_{t2}$ represent the distance between the focal points and the trap radii of the left and right beams, respectively. The trap radius is the distance that indicates how far the center of the trapped particle can depart from the focal point. Then, the PMF can be calculated from the following equation:

$$u(r) = -k_{\mathrm{B}}T\ln p(r) - k_{\mathrm{B}}T\ln Z_r + k_{\mathrm{B}}T\ln\left(r\int_0^{r_{\mathrm{t1}}}\int_0^{2\pi}\int_0^{2\pi}\exp\left(-\frac{v_1(a_1,r_1)+v_2(a_2,r_2)}{k_B T}\right)H_1(r_{\mathrm{t2}}-r_2)d\theta_2 d\theta_1 dr_1\right), \quad (40)$$

where $v_i$ ($i$ = 1 or 2) is the parabolic-type potential, the expression of which is $v_i = a_i r_i^2 - \varepsilon_i k_{\mathrm{B}}T$. $H_1$ is the Heaviside step function containing a property $H_1(0) = 1$. The distance between the right focal point and the center of the right particle $r_2$ is expressed as

$$r_2^2 = l_{\mathrm{t}}^2 + r^2 + r_1^2 - 2l_{\mathrm{t}}r_1\cos\theta_1 + 2rr_1\cos(\theta_1-\theta_2) - 2l_{\mathrm{t}}r\cos\theta_2. \quad (41)$$

In the 2D-DPPR method, Eq. (40), $p(r)$ measured by the dual-beam optical tweezers, and the absolute temperature are used to obtain $u(r)$. Although there are five unknown parameters ($a_1$, $a_2$, $\varepsilon_1$, $\varepsilon_2$, and $Z_r$) in Eq. (40), the concrete values of the latter three parameters ($\varepsilon_1$, $\varepsilon_2$, and $Z_r$) are not needed, because $u(r)$ at the long distance is equal to zero (i.e., use of the boundary condition). The remaining parameters ($a_1$ and $a_2$) can be roughly estimated from the videos measured by the single-beam optical tweezers, and determined by using the boundary condition, $du(r)/dr = 0$. As a supplementary note, we write that the shape of the trap potentials ($v_1$ and $v_2$) do not necessarily have to be parabolic in order to calculate the PMF, as long as they are radially symmetric.

In summary, we have created the calculation methods of the PMF between colloidal particles from the probability density functions measured with the laser tweezers. Derivation processes of the 1D-WPR, 1D-PPR, 2D-PPR, and 2D-DPPR methods are explained. These methods are important techniques for more accurate measurements of the PMF. For example, Amano *et al*. [6] obtained the density distribution of the smaller colloidal particles near the surface of the larger colloidal particle by using experimental data from LOT [1]. Such a density distribution can be obtained more accurately if the 1D-PPR method is used. The 2D-PPR method can be used for study of three-body interaction. This is because it can calculate the PMF between two colloidal particles near a substrate surface. We will show the demonstration results of the 1D-WPR, 1D-PPR, 2D-PPR, and 2D-DPPR methods by using the experimental data in the future.

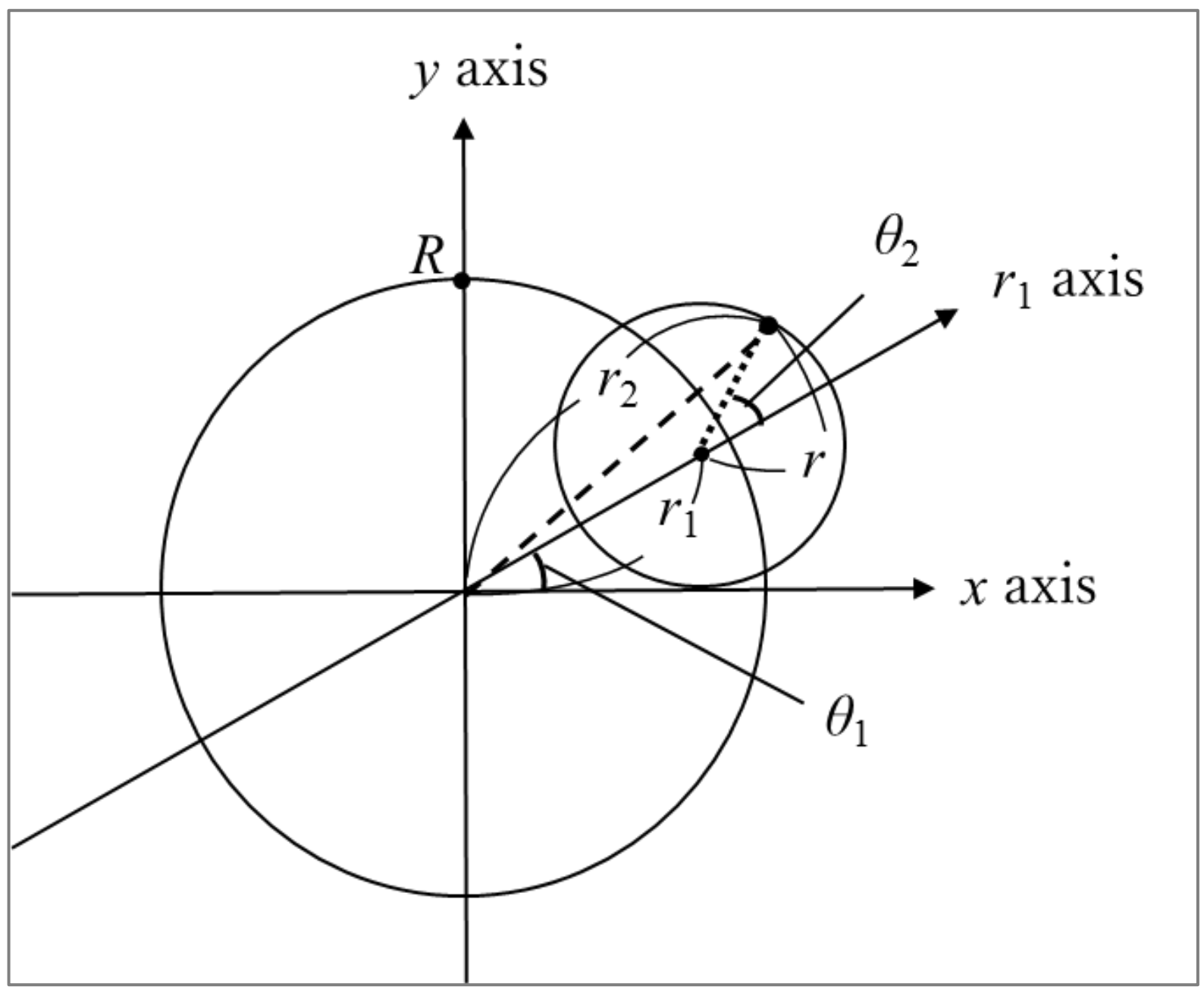

Fig. 1 Schematic of the polar coordinates. $x$ and $y$ axes are fixed, but $r_1$ axis rotates according to $\theta_1$. Two colloidal particles are constrained within the circle with radius $R$.

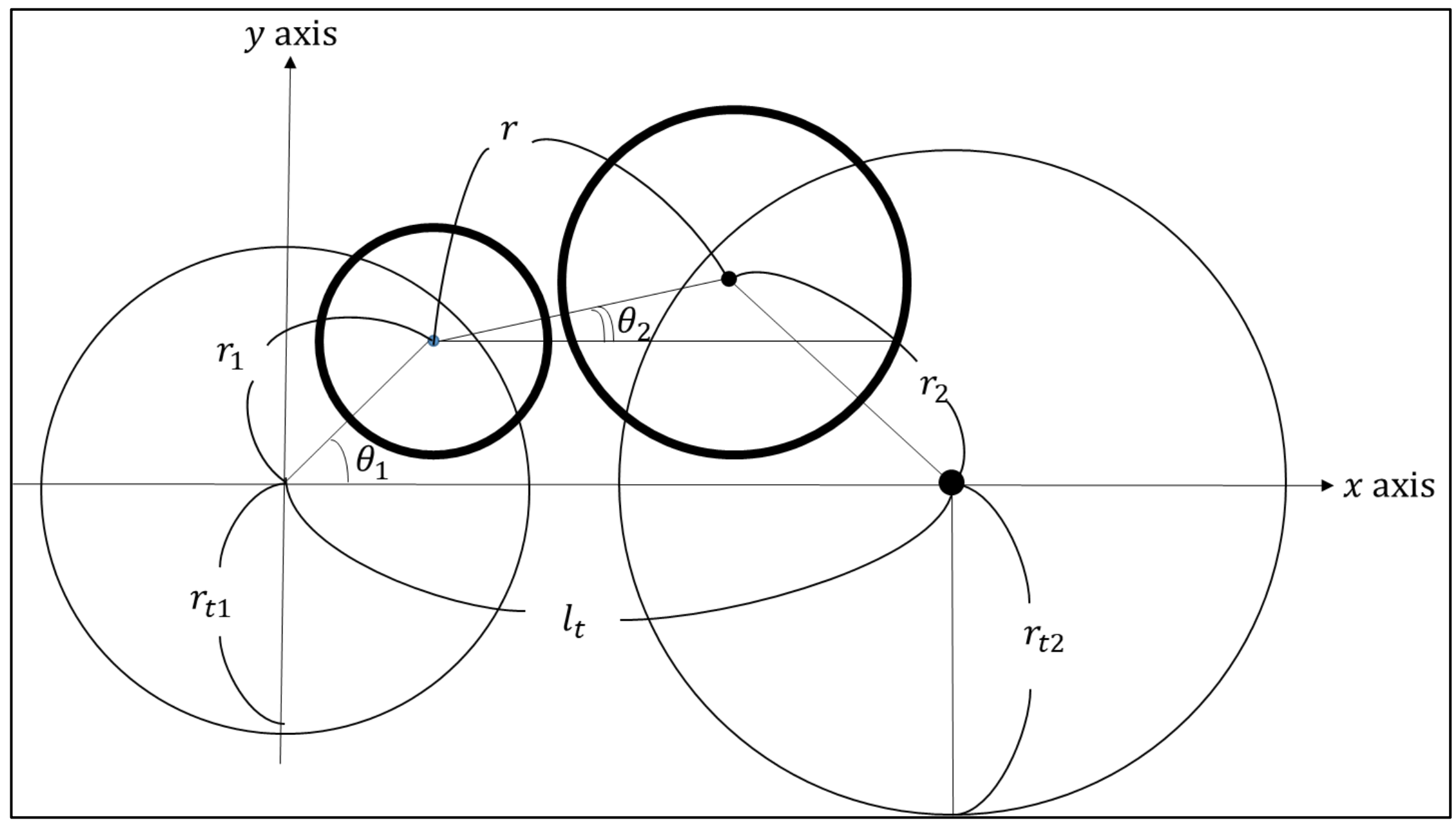

Fig. 2 Schematic of the polar coordinates, where the trapped colloidal particles are drawn by bold lines. Centers of the left and right colloidal particles are constrained within the circles with radii $r_{t1}$ *and* $r_{t2}$, respectively.

## ACKNOWLEDGEMENTS

We thank M. Iwaki, M. Maebayashi, T. Sakka, T. Ishihara, K. Hashimoto, H. Kitaoka, and M. Narita for useful discussions and helpful supports. This work was financially supported by the Research Institute of Meijo University.